\documentclass[11pt]{article}
\usepackage{moriond,epsfig}

\def\Journal#1#2#3#4{{#1} {\bf #2}, #3 (#4)}

\def\PRD{{\em Phys. Rev.} D}

\def\be{\begin{equation}}
\def\bea{\begin{eqnarray}}

\def\bsmudec{$B_{s}^{0} \rightarrow {D}_s^{-} \mu^+ X \hspace{1mm}$}

\newcommand{\dzer}         {D\O}
\newcommand{\ds}        {\ensuremath{{D_{s}^{-}}}}

\begin{document}
\vspace*{4cm}
\title{$B^0_s$ MIXING STUDIES AT THE TEVATRON}

\author{MD. NAIMUDDIN (on Behalf of CDF and D\O\ collaborations)  }

\address{Department of Physics and Astrophysics, University of Delhi, Delhi 110007, India}

\maketitle\abstracts{Measurement of the $B^0_s$ oscillation frequency via $B^0_s$ mixing analysis provides a powerful constraint on CKM matrix elements. This note briefly reviews the motivation behind these analyses and describes the various steps that go into a mixing measurement. Recent results on $B^0_s$ mixing obtained by the CDF and D\O\ collaborations using the data samples collected at Tevtron Collider in the period 2002 - 2005 are presented.}

\section{Introduction}
\label{intro}
Mixing is the process whereby some neutral mesons change from their particle to their anti-particle state, and vice versa. This kind of oscillation of flavor eigenstates into one another was first observed in the $K^{0}$ meson system. It has since then been seen for $B$ mesons, first in an admixture of $B^0_d$ and $B^0_s$ by UA1\cite{ua1} and then in $B^0_d$ mesons by ARGUS \cite{argus}. The combinations of these results already indicated that the frequency of $B^0_s$ mixing oscillations was higher than the frequency of $B^0_d$ oscillation. The frequency of the oscillation is proportional to the  small difference in mass between the two eigenstates, $\Delta m$,  and for the $B^0_d - \bar{B^0_d}$ system can be translated into a measurement of the CKM element $|V_{td}|$.  $\Delta m_d$ has been precisely measured (the world average is $\Delta m_d=0.502 \pm 0.007$ ps$^{-1}$) \cite{pdg} but large theoretical uncertainties dominate the extraction of $|V_{td}|$ from $\Delta m_d$. This problem can be reduced if the $B_s^0$ mass difference, $\Delta m_s$, is also measured. $|V_{td}|$  can then be extracted with better precision from the ratio:
\begin{equation}
\frac{\Delta m_s}{\Delta m_d}=\frac{m(B_s^0)}{m(B_d^0)}\xi ^2 {\mid \frac{ V_{ts
}}{V_{td}}\mid}^2
\end{equation}
where $\xi$ is estimated from Lattice QCD calculations to be $1.15 \pm 0.05_{-0.00}^{+0.12}$ \cite{pdg}. The above has motivated many experiments to search for $B_s^0$ oscillations though a statistically significant signal hasn't been observed before this work, a lower limit of $\Delta m_s >16.1 $ ps$^{-1}$ at 95\% C.L.\cite{pdg} has been set. Since this current limit indicates that the $B_s^0$ oscillations are at least 30 times faster than the $B_d^0$ oscillations, a $B_s^0$ mixing measurement is experimentally very challenging. If the Standard Model is correct, then $\Delta m_s$ is expected from global fits to the unitarity triangle to be in the range $(16.2 - 24.5)$ ps$^{-1}$ at the one standard deviation confidence level \cite{Charles:2004jd}. \\
In the $B^0_s$-$\bar{B}^0_s$ system there are two
mass eigenstates, the heavier (lighter) one having mass
$M_H$ ($M_L$) and decay width $\Gamma_H$ ($\Gamma_L$). 
Denoting $\Delta m_s = M_H - M_L$ and 
$\Delta \Gamma_s = \Gamma_L - \Gamma_H$, the time dependent
probability  that a $B^0_s$ oscillates into a $\bar{B}^0_s$
(or vice versa) is given by 
$P^{\mathrm{osc}} = \Gamma e^{-\Gamma t}(1 - \cos\Delta m_s t)/2$
while the probability that the $B^0_s$ does not oscillate is
given by
$P^{\mathrm{nos}} = \Gamma e^{-\Gamma t}(1 + \cos\Delta m_s t)/2$, 
assuming that $\Delta\Gamma_s$ is small and neglecting CP violation.

\section{Tevatron Detectors}
\subsection{CDF detector}
\label{cdfdetector}
The CDF detector is described in detail elsewhere~\cite{cdfdet}.
The components most relevant to this analysis are briefly described here. 
The tracking system is in a 1.4 T axial magnetic field and consists of a silicon 
microstrip detector surrounded by an open-cell wire drift chamber (COT). 
The muon detectors used for this analysis
 are the central muon drift chambers (CMU), covering the pseudorapidity range 
$|\eta|<0.6$, and the extension muon drift chambers (CMX), covering $0.6<|\eta|<1.0$, 
where $\eta = - \ln [(\tan(\theta/2)]$ and $\theta$ is the polar angle.

\subsection{D\O\ detector}
\label{d0detector}
The D\O\ detector is described in detail elsewhere ~\cite{run2det}.
The central tracking and muon systems are the components
most important to this analysis.
The central tracking system consists of a
silicon microstrip tracker (SMT) and a central fiber tracker (CFT),
both located within a 2~T superconducting solenoidal
magnet, with designs optimized for tracking and
vertexing for pseudorapidities $|\eta|<3$ and $|\eta|<2.5$, respectively.
An outer muon system, at $|\eta|<2$,
consists of a layer of tracking detectors and scintillation trigger
counters in front of 1.8~T toroids, followed by two similar layers
after the toroids ~\cite{run2muon}.

\section{Analysis Technique}
The analysis starts with the reconstruction of the final state of the $B^0_s$ meson. At CDF, the $B^0_s$ mesons reconstructed in semileptonic as well as in hadronic decays of $B^0_s$. D\O\ uses only the semileptonic decays of the $B^0_s$ meson for the final state reconstruction. CDF has about 18,200 $B^0_s \to \ds ( \to \phi \pi) \mu^+ X$ in 765 $pb^{-1}$ of data. It also has about 2300 $B^0_s$ signal candidates in Hadronic channels.
D\O\ analyzes 1 fb$^{-1}$ of data and reconstruct $26,710 \pm 560$ events in the decay $B^0_s \to \ds ( \to \phi \pi) \mu^+ X$.

In order to know the initial flavor of the $B^0_s$ mesons, an Initial State Flavor tagging technique is used. The second $B$ meson (or baryon) in the event was used to tag the initial flavor of the reconstructed $B^0$ meson. The tagging technique utilized information from identified leptons (muons and electrons) and reconstructed secondary vertices. For reconstructed \bsmudec decays both leptons having the same sign would indicate that one $B$ hadron had oscillated while opposite signs would indicate that neither (or both) had oscillated. The performance of the flavor tagging is characterized by the efficiency, $\epsilon = N_{\mathrm{tag}}/N_{\mathrm{tot}}$, where $N_{\mathrm{tag}}$ is the number of tagged $B^0_s$ mesons, and $N_{\mathrm{tot}}$ is the total number; the tag purity $\eta_s$, defined as $\eta_s = N_{\mathrm{cor}}/N_{\mathrm{tag}}$, where $N_{\mathrm{cor}}$ is the number of $B^0_s$ mesons with correct flavor identification; and dilution, related to purity as $\mathcal{D} = 2\eta_s -1$. The tagging can be performed on the opposite side as well as on the same side of the reconstructed $B^0_s$ meson. Three main tagging algorithms were used in the present analysis viz. Soft Lepton Tagging (where lepton could be a muon or an electron), Jet Charge Tagging and Same Side Tagging (only at CDF). The performance of the CDF's combined Opposite Side Tagging (OST) is, $\epsilon D^2 = (1.55 \pm 0.020 \pm 0.014)\%$. The Same Side Tagging (SST) Performance is, $\epsilon D^2 = (4.0+0.9-1.2)\%$. The D\O\ combined OST Performance is, $\epsilon D^2 = (2.48 \pm 0.21 +0.08 -0.06)\%$. The taggers were tuned by measuring the $B^0_d$ mixing oscillations and D\O\ finds $\Delta m_{d} = 0.506 \pm 0.020 \pm 0.014$ ps$^{-1}$ in good agreement with the world average of $\Delta m_{d} = 0.502 \pm 0.007$ ps$^{-1}$ \cite{pdg}. After applying the tagging to the D\O\ data, $5601 \pm 102$ tagged events were found.

Once the tagging is performed, the proper decay time of candidates is needed. The proper decay length of each $B^0_s$ mesons is found as $c t_{B^0_s} = (\vec{L}_{T} \cdot M_{B^0_s}/(p_{T}^{B^0_s})$, where $\vec{L}_T$ is the vector in the transverse plane from the primary to the $B^0_s$ decay vertex, and $M_{B^0_s} = 5.3696$~GeV~\cite{pdg}.  However, in the case of semileptonic $B^0_s$ decay, the undetected neutrino does not allow a precise determination of the meson's momentum and Lorentz boost. To take into account the effects of neutrinos and other lost or non-reconstructed particles, a $K$ factor was estimated from Monte Carlo (MC) simulation by finding the distribution of $K = p_T(\mu D_s)/p_T(B)$ for a given decay channel. The proper decay length of each $B^0_s$ meson is then $c t_{B^0_s} = l_M \cdot K$, where $l_M = (\vec{L}_{T}\cdot \vec{p}_T^{\mu \ds})/(p_{T}^{\mu \ds})^{2} \cdot M_{B^0_s}$ is the measured visible proper decay length (VPDL). 
  The VPDL uncertainty was determined by the vertex fitting procedure and track parameter uncertainties. To account for possible mismodeling of detector uncertainties, resolution scale factors were introduced as determined by examining the pull distribution of the the vertex positions of a sample of $J/\psi \rightarrow \mu \mu$ decays.

\section{Results and Conclusions}
\begin{figure}[hbtp]
\centerline{\epsfxsize 3.0 truein
\epsfbox{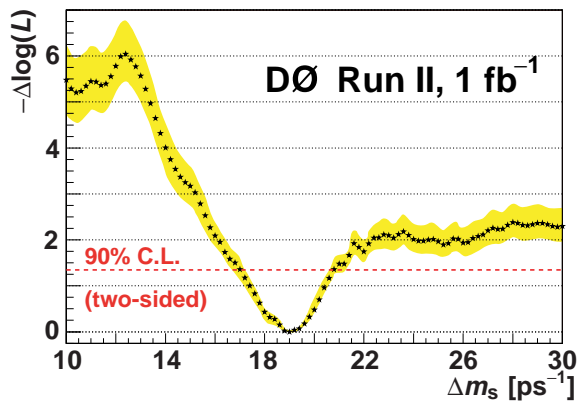}}
\caption{Likelihood Scan plot for D\O\ data. Yellow band shows the effect of systematics.}
\label{lhood}
\end{figure}

Using the Amplitude Fit Method \cite{ampfit} and 365 pb $^{-1}$ of data, CDF puts a limit on $B^0_s$ oscillations frequency of 8.6 ps$^{-1}$ and sensitivity of 13.0 ps$^{-1}$ at 95\% C.L.
\begin{figure}[htb]
\begin{center}
\includegraphics[scale=0.3]{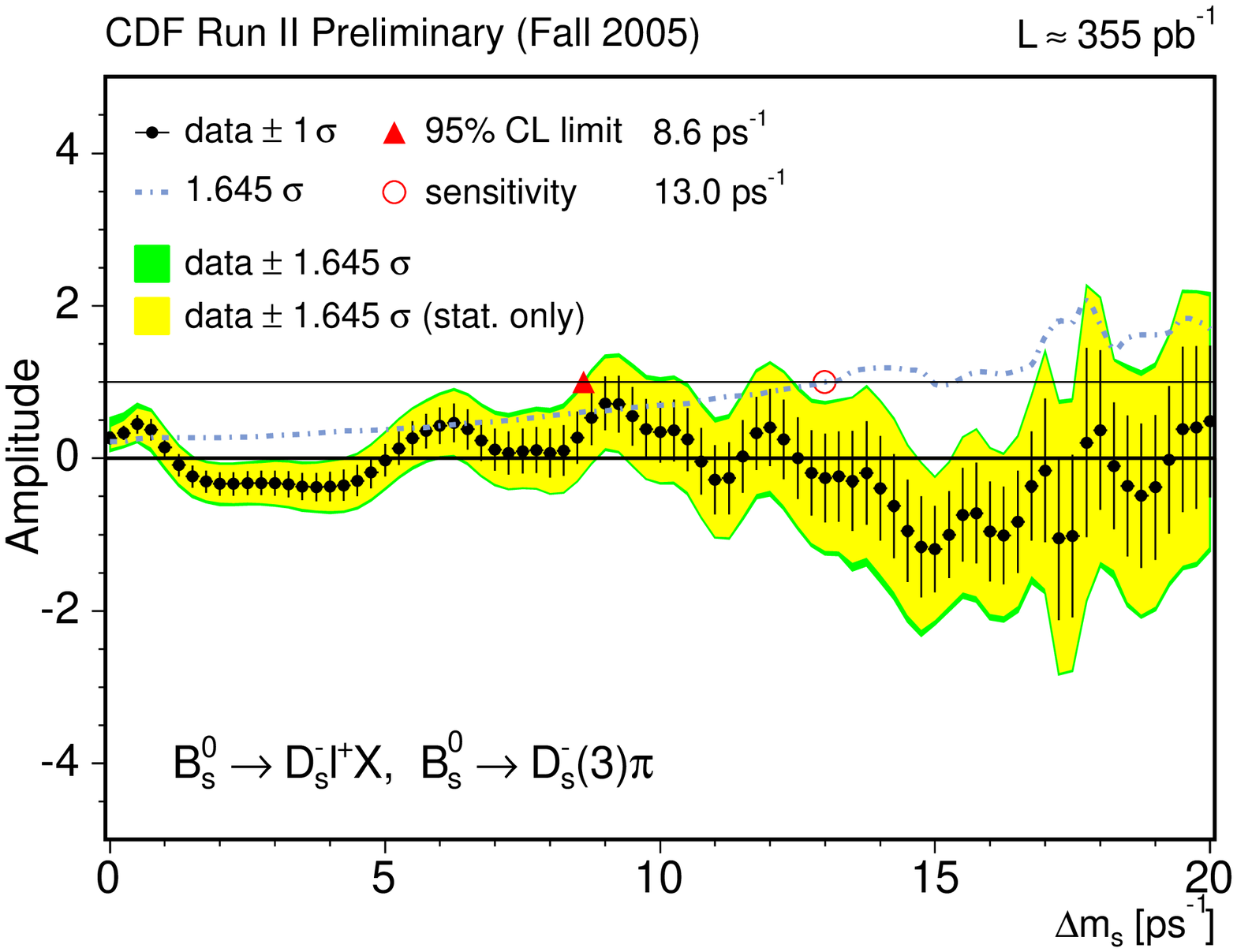}
\includegraphics[scale=0.4]{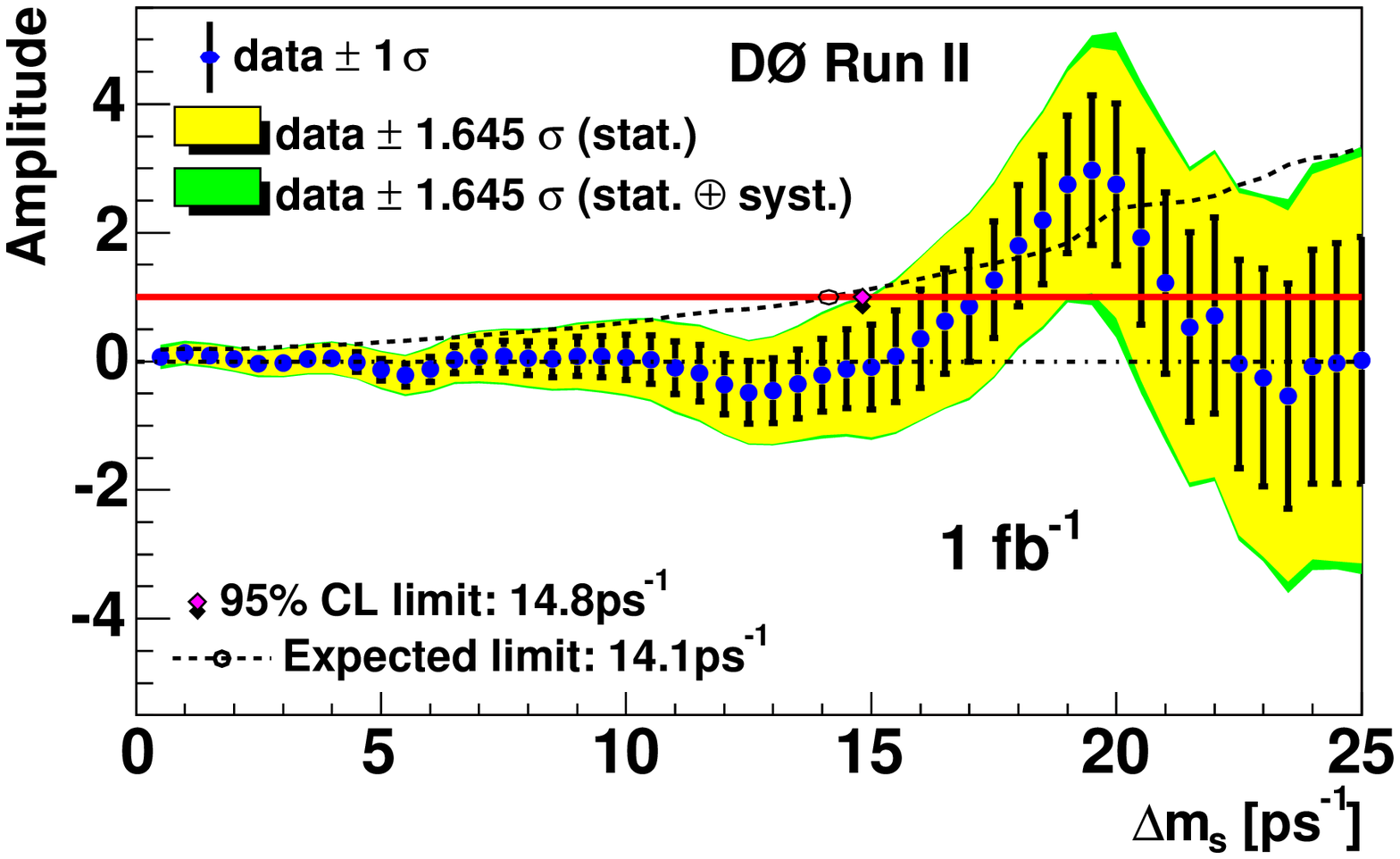}
\caption[]{Amplitude fit Scan for CDF (left) and D\O\ (right) data.}
    \label{ampl}  
\end{center}
\end{figure}
D\O\ using the similar method finds a limit of 14.8 ps$^{-1}$ and sensitivity of 14.1 ps$^{-1}$ at 95\% C.L. D\O\ also performed a likelihood scan as a function of $\Delta m_s$. Figure~\ref{lhood} shows the value of $-\Delta \log\mathcal{L}$ as a function of $\Delta m_s$, indicating a favored value of 19~ps$^{-1}$, while variation of $\log\mathcal{L}$ from the minimum indicates an oscillation frequency of $17.0 < \Delta m_s < 21.0$~ps$^{-1}$ at the 90\% C.L. The uncertainties are approximately Gaussian inside this interval. The parametrized MC test shows that for a true value of $\Delta m_s = 19$~ps$^{-1}$, the probability was 15\% for measuring a value in the range $17.0 < \Delta m_s < 21.0$~ps$^{-1}$ with a -$\Delta \log\mathcal{L}$ lower by at least 1.9 than the corresponding value at $\Delta m_s = 25 ps^{-1}$. \cite{d0final} \\
   To test the statistical significance of the observed minimum, an ensemble test using the data sample was performed by randomizing the flavor tag and retaining all other information for the candidate, effectively simulating a $B^0_s$ oscillation with an infinite frequency. The ensemble test results shows that the probability to observe a minimum in the range $16.0 < \Delta m_s < 22.0$~ps$^{-1}$ with a decrease in $-\log\mathcal{L}$ with respect to the corresponding value at $\Delta m_s = 25$ ps$^{-1}$ of more than 1.7, corresponding to our observation including systematic uncertainties, was found to be $(5.0 \pm 0.3)\%$. This range of $\Delta m_s$ was chosen to encompass the world average lower limit and the edge of our sensitive region.

\section{outlook}
Further improvements are planned for future which includes improvements both in detector and analysis technique. After this conference, CDF updated their $B^0_s$ mixing analysis and the latest CDF results can be found here~\cite{cdfresult}.

\section*{Acknowledgments}
I am grateful to Prof. D.S. Kulshrestha for his continuous guidance and helpful suggestions. My sincere  thanks to Fermilab and the organizers of the Moriond QCD conference for supporting me to attend this conference.

\end{document}